\documentclass[11pt]{article}
\newcommand{\Comment}[1]{{}}
\usepackage{amsfonts,amsthm,amsmath,amssymb,slashed}
\usepackage[textwidth = 430 pt, textheight = 630 pt]{geometry}
\usepackage{color}

\Comment{\usepackage{color}
\definecolor{MyDarkBlue}{rgb}{0.15,0.15,0.45}
\usepackage[linktocpage=true]{hyperref}
\hypersetup{
colorlinks=true,
citecolor=MyDarkBlue,
linkcolor=MyDarkBlue,
urlcolor=MyDarkBlue,
pdfauthor={Horatiu Nastase and Jacob Sonnenschein},
pdftitle={Euler fluid as a gauge theory, and an action for the Euler fluid},
pdfsubject={hep-th}
}

\usepackage[numbers,sort&compress]{natbib}
\usepackage{hypernat}}
\usepackage{cite}

\newcommand\ignore[1]{}
\def\one{{\,\hbox{1\kern-.8mm l}}}

\def\Tr{{\rm Tr\, }}

\def\a{\alpha}

\def\d{\partial}

\def\Tr{\mathop{\rm Tr}\nolimits}

\newcommand{\Cset}{{\,\,{{{^{_{\pmb{\mid}}}}\kern-.45em{\mathrm C}}}}}

\newcommand{\be}{\begin{equation}}
\newcommand{\bea}{\begin{eqnarray}}

\newcommand{\ee}{\end{equation}}
\newcommand{\eea}{\end{eqnarray}}

\newcommand{\Vol}{\text{Vol}}
\newcommand{\zb}{\bar{z}}
\newcommand{\wb}{\bar{w}}

\newcommand{\hpsi}{\hat{\psi}}
\newcommand{\htheta}{\hat{\theta}}
\newcommand{\hphi}{\hat{\phi}}

\newcommand{\homega}{\hat{\omega}}
\newcommand{\tomega}{\tilde{\omega}}
\newcommand{\tm}{\tilde{m}}

\begin{document}

\renewcommand{\thefootnote}{\fnsymbol{footnote}}

\makeatletter
\@addtoreset{equation}{section}
\makeatother
\renewcommand{\theequation}{\thesection.\arabic{equation}}

\rightline{}
\rightline{}




\begin{center}
{\LARGE \bf{\sc Penrose limits of I-branes, twist-compactified D5-branes and spin chains}}
\end{center} 
 \vspace{1truecm}
\thispagestyle{empty} \centerline{
{\large \bf {\sc Marcelo Barbosa${}^{a}$}}\footnote{E-mail address: \Comment{\href{mailto:mr.barbosa@unesp.br}}
{\tt mr.barbosa@unesp.br}},
{\large \bf {\sc Horatiu Nastase${}^{a}$}}\footnote{E-mail address: \Comment{\href{mailto:horatiu.nastase@unesp.br}}
{\tt horatiu.nastase@unesp.br}},
{\large \bf {\sc Carlos Nunez${}^{b}$}}\footnote{E-mail address: \Comment{\href{mailto:c.nunez@swansea.ac.uk}}
{\tt c.nunez@swansea.ac.uk}}
{\bf{\sc and}}
{\large \bf {\sc Ricardo Stuardo${}^{b}$}}\footnote{E-mail address: \Comment{\href{mailto:ricardostuardotroncoso@gmail.com}}
{\tt ricardostuardotroncoso@gmail.com}}
                                                        }

\vspace{.5cm}


\centerline{{\it ${}^a$Instituto de F\'{i}sica Te\'{o}rica, UNESP-Universidade Estadual Paulista}} 
\centerline{{\it R. Dr. Bento T. Ferraz 271, Bl. II, Sao Paulo 01140-070, SP, Brazil}}
\vspace{.3cm}
\centerline{{\it ${}^b$Department of Physics, Swansea University,}}
\centerline{{\it Swansea SA2 8PP, United Kingdom }} 
\vspace{1truecm}

\thispagestyle{empty}

\centerline{\sc Abstract}

\vspace{.4truecm}

\begin{center}
\begin{minipage}[c]{380pt}
{\noindent In this paper we consider the Penrose limit in the case of two  gravity duals. One of them, consists of compactified I-branes (intersecting 
sets of D5-branes over $(1+1)$ dimensions). The second 
consists of D5-branes compactified on a circle. Both compactifications preserve SUSY. We find a match of the oscillators and masses of string modes on the resulting 
pp wave against a spin chain in a $(2+1)-$dimensional field theory in the first case, and a spin chain in a $(1+1)-$dimensional
field theory in the second.
}
\end{minipage}
\end{center}

\vspace{.5cm}

\setcounter{page}{0}
\setcounter{tocdepth}{2}

\newpage

\tableofcontents
\renewcommand{\thefootnote}{\arabic{footnote}}
\setcounter{footnote}{0}

\linespread{1.1}
\parskip 4pt



\section{Introduction}

In the AdS/CFT correspondence \cite{Maldacena:1997re}, and even more so in the gauge/gravity duality generalizations, often times 
the holographic dictionary and/or the exact matching of the two sides is not very clear.  
In order to gain additional information, through specializing to a useful subset, one tool at our disposal is the Penrose limit.
The Penrose limit leads, on the gravity side, to a plane-parallel wave (pp-wave), and on the field theory side, to a spin chain, as first seen in  
\cite{Berenstein:2002jq}. The Penrose limit  has been used to make progress in less understood cases, for instance in 
\cite{Araujo:2017hvi,Itsios:2017nou,Nastase:2021qvv,Nastase:2022dvy,Barbosa:2023vmy}.

A particular case of interest is that of confining field theories.
In a confining field theory, like the $(3+1)-$dimensional case of \cite{Maldacena:2000yy} or the $(2+1)-$dimensional 
case of MNa \cite{Maldacena:2001pb}, the Penrose limit leads to a theory of long cyclic hadron-like objects, dubbed 
"annulons" in \cite{Gimon:2002nr}. But the resulting "spin chain" is not well understood either, though in the MNa case 
progress was made in \cite{Nastase:2021qvv}. 

It would then be useful to study confining theories in different dimensions, like the (fibered) I-brane case, whose (singular) holographic dual was 
written in  \cite{Itzhaki:2005tu}, and whose nonsingular supergravity background was found and studied 
in \cite{Nunez:2023nnl}. Also, in  the related case of $S^1$ twisted-compactified D5-branes whose nonsingular background was found and studied in 
\cite{Nunez:2023xgl}. This is the subject of this paper. The I-brane theory is localized at the $(1+1)-$dimensional 
intersection of two sets of D5-branes, but it was found in  \cite{Itzhaki:2005tu} that one dimension appears dynamically, 
so the theory should be understood in $(2+1)$ dimensions.
On the other hand, the theory of D5's compactified on $S^1$ is dual to a confining $(4+1)$ dimensional QFT with eight supercharges.

The paper is organized as follows.
In Section \ref{sectionIbranes} we discuss in detail the Penrose limit  along two possible geodesics in the non-singular geometry of I-branes \cite{Nunez:2023nnl}. The two geodesics lead to the same pp-wave and, after some choice is made, we write it in the form of a parallelizable plane wave \cite{Sadri:2003ib}. We then discuss the spectrum of the string and the associated spin chain, matching oscillations in QFT and in string theory.

Analogously, in Section \ref{sectiontwistedD5} we discuss the Penrose limit in the background of $S^1$-twisted compactified D5 branes (preserving eight SUSYs) \cite{Nunez:2023xgl}. The pp-wave is also parallelizable and preserves 24 SUSYs. A proposal for the associated spin chain is given.
Section \ref{discusion-conclusion} presents a summary and closing remarks, together with the proposal for further study in different backgrounds.

\section{Penrose limit of fibered I-branes and dual spin chain}\label{sectionIbranes}

The $(1+1)-$dimensional theory obtained when two stacks of D5-branes intersect along two space-time directions, is called I-brane theory \cite{Itzhaki:2005tu}. As explained there,
an extra worldvolume coordinate appears, leading to a $(2+1)-$dimensional theory. After a twisted compactification on a shrinking circle,  
the dual background was found to be {\em nonsingular} in \cite{Nunez:2023nnl}. This provided an 'IR completion' 
to an otherwise singular gravity dual.\footnote{See \cite{Nayak:2010bw} for a Penrose limit on the singular background.}

The supergravity background dual to the fibered I-branes, in the D5-brane string frame is relevant to the dual field theory.
For simplicity, we  write the S-dual NS5-brane theory, keeping in mind that for the field theory analysis we should work with D5-branes. The NS5 solution reads
\bea
        ds^{2}_{st} &=&-dt^{2}+dx^{2}+ 4Q^{2}f(\rho)d\varphi^{2} + \frac{d\rho^{2}}{f(\rho)} 
        +\frac{N_{B}}{4}\left[\homega_{1}^{2}+\homega_{2}^{2}
        +\left( \homega_{3}-e_{A}A \right) ^{2}\right] \cr
&&        +\frac{N_{A}}{4}\left[ \tomega_{1}^{2}+\tomega_{2}^{2}
        +\left( \tomega_{3}^{2}-e_{B}B \right) ^{2}\right] \cr
        H_{3} &=& \frac{2}{e_{A}}d(\homega_{3}\wedge A)  
        +\frac{2}{e_{B}}d(\tomega_{3}\wedge B)  
            + 2N_{B} \Vol(S^{3}_{A}) + 2N_{A} \Vol(S^{3}_{B}).\cr
        \Phi &=& -Q\rho,\label{diego}
    \eea
where 
$e^{2}_{A,B} = 8/N_{B,A}$,  $N_A, N_B$ are integers, $\homega^{i}$ and $\tomega^{i}$ are the Maurer-Cartan forms for $su(2)$, given by
    \begin{eqnarray}
& & \hat{\omega}_1= \cos {\psi_A} d{\theta_A} +\sin{\psi_A}\sin\theta_A d{\phi_A},\;\;\;\;\; \tilde{\omega}_1= \cos {\psi_B} d {\theta_B} +\sin {\psi_B}\sin\theta_B d{\phi_B},\nonumber\\
& & \hat{\omega}_2= -\sin{\psi_A} d{\theta_A} +\cos {\psi_A}\sin\theta_A d{\phi_A},\;\;\tilde{\omega}_2= -\sin {\psi_B} d{\theta_B} +\cos {\psi_B} \sin\theta_B d {\phi_B},\nonumber\\
& & \hat{\omega}_3= d{\psi_A} +\cos{\theta_A}d{\phi_A},\;\;\;\;\;\;\;\; ~~~~~~~~~\;\;\;\;\; \tilde{\omega}_3= d{\psi_B} +\cos {\theta_B}d {\phi}_B.\nonumber
    \end{eqnarray}
The 1-forms are
    \begin{equation}
    \begin{aligned}
         A &= Q_{A}\zeta(\rho)d\varphi, \quad 
         B = Q_{B} \zeta(\rho) d\varphi,
    \end{aligned}
    \end{equation}
where we have defined the functions
    \begin{equation}
        f(\rho) = 1 - \tm\, e^{-2Q\rho} - \frac{(Q^{2}_{A}+Q^{2}_{B})}{2Q^{2}} e^{-4Q\rho}, \quad
        \zeta(\rho) = e^{-2Q\rho} - e^{-2Q\rho_{+}}\;,\label{leo}
    \end{equation}
with $Q=\sqrt{\frac{1}{N_A}+\frac{1}{N_B}}$ is a background charge in the string worldsheet theory, 
and $\rho_+$ is the larger of the two solutions of $f(\rho)=0$, $\rho_\pm$. We also defined $\tilde m=\frac{m}{4Q^2}$, with $m$ the standard mass 
parameter.

We note that $\rho=\rho_+$ is the end of the space, corresponding to the IR of the field theory. 

\subsection{Penrose limit}
Below we study the Penrose limit for the geometry in eqs.(\ref{diego})-(\ref{leo}). We find two possible geodesics about which to expand and write the associated pp-wave. We then show the equivalence of the two plane waves obtained. After that we comment on the amount of SUSY preserved and quantize the string on this background.

\subsubsection{First limit: geodesic on $\psi_A,\psi_B$}

We consider a geodesic moving (besides the time $t$) in the directions $\psi_A$ and $\psi_B$, and fixed at $x=\phi_A=\phi_B=0$, 
(we could have replaced $x=0$ with some arbitrary $x=x_0$ as well), $\theta_A=\theta_B=\pi/2$ and $\rho=\rho_+$. 

Note that $\phi_A=\phi_B=0$ and $\theta_A=\theta_B=\pi/2$ are necessary in order to have a solution of the geodesic equation
$\frac{dx^\mu}{d\lambda}+\Gamma^{\mu}_{\nu\rho}\frac{dx^\nu}{d\lambda}\frac{dx^\rho}{d\lambda}=0$. 
Moreover, as we are interested in understanding the IR of the theory, we must focus on the geodesics that sit at $\rho=\rho_+$. 

The geodesics we are interested in, are further defined by a rotation of an angle $\a$ between the two spatial directions 
$\psi_A$ and $\psi_B$. This means that we make the change of variables
\bea
\psi_{A} &\rightarrow& \frac{2}{\sqrt{N_{B}}} 
                \left( \cos(\alpha)\psi_{A} + \sin(\alpha)\psi_{B}\right)\equiv \tilde \psi_A,\cr
        \psi_{B} &\rightarrow& \frac{2}{\sqrt{N_{A}}}
                \left( -\sin(\alpha)\psi_{A} + \cos(\alpha)\psi_{B}\right)\equiv \tilde \psi_B\;,\label{rota1}
\eea
and take the geodesic to be on $\tilde \psi_B$. 
One defines lightcone coordinates as usual, 
\bea
t &=& \frac{1}{\sqrt{2}}(u-v),\cr
\tilde \psi_{A} &=& \frac{1}{\sqrt{2}}(u+v).\label{light1}
\eea

The coordinate $\rho$ is kept always slightly off the special point $\rho=\rho_+$, where 
$f(\rho_+)=0$, by the rescaling with $L$ written below. The Penrose rescaling is 
\bea
&&v\rightarrow \frac{v}{L^{2}}, \quad 
        \theta_{A,B} \rightarrow \frac{\pi}{2}+\frac{2\theta_{A,B}}{L\sqrt{N_{B,A}}}, \quad
        \phi_{A,B} \rightarrow \frac{2\phi_{A,B}}{L\sqrt{N_{B,A}}}, \quad
        x \rightarrow \frac{x}{L}, \cr
&&\rho \rightarrow \rho_{+} + \frac{\rho}{L^2}, \quad
\psi_{B} \rightarrow \frac{\psi_{B}}{L}\;, \quad u\rightarrow u\;,\quad \varphi\rightarrow \varphi.
\eea
Note that the $\varphi$ was not rescaled, since it will become an angular variable in the pp wave (so it is not a length variable, 
as to be rescaled by $1/L$), in the same way that it was done, for instance, in \cite{Nastase:2021qvv,Barbosa:2023vmy}. We also need to change coordinates according to \footnote{ Note that it is only the redefined $\rho$ that rescales by $1/L$ in the Penrose limit, as required by the Penrose theorem, since only this variable becomes the radial coordinate on the plane.}
\begin{equation}
  \rho\to \frac{\sinh( Q(\rho_+ - \rho_-))}{e^{Q(\rho_+ -\rho_-)}} \rho^2, ~~~\varphi\to \frac{e^{Q(\rho_+- \rho_-)}}{4 Q^2 \sinh(Q(\rho_+ -\rho_-))}\varphi. \label{rescalingwave1} 
\end{equation}

Then in the $L\rightarrow\infty$ limit, after rescaling the metric by $L^2$, we obtain the pp wave 
\bea
L^{2} ds^{2} &=& 2du\left(dv + A_{\varphi}d\varphi - A d\phi_{A} - B d\phi_{B} \right) + dx^{2} + d\rho^{2} + \rho^{2}d\varphi^{2}\cr
&& + d\theta^{2}_{A} + d\phi^{2}_{A}+ d\theta^{2}_{B} + d\phi^{2}_{B} + d\psi^{2}_{B}, \cr
L^{2} H_{3} &=& dA\wedge d\phi_{A} \wedge du + dB\wedge d\phi_{B} \wedge du + dA_{\varphi} \wedge d\varphi\wedge du,\cr
\Phi &=& 0\;.
\eea
We have defined 
\bea
A_{\varphi} &=& \frac{1}{2} e^{-2Q\rho_{+}}  \rho^{2}  \Big( Q_{A}\cos(\alpha) -Q_{B}\sin(\alpha) \Big), \cr
A &=& \sqrt{\frac{2}{N_{B}}} \cos(\alpha) \theta_{A}, \quad 
B = \sqrt{\frac{2}{N_{A}}} \sin(\alpha) \theta_{B}.\label{diego10}
\eea

We can write the B-field corresponding to the above field strength $H_3$, 
\be
L^{2} B_{2} = A d\phi_{A} \wedge du + B d\phi_{B} \wedge du 
                + A_{\varphi} d\varphi \wedge du.
\ee

Finally, we go to Cartesian coordinates, from the $(\rho,\varphi)$ space to $(x_1,x_2)$, and define the corresponding
rotated $A_\varphi$ into $(A_1,A_2)$, 
\bea
A_{1} &=& - \frac{1}{2}x_{2} e^{-2Q\rho_{+}}  \left( Q_{A}\cos(\alpha) -Q_{B}\sin(\alpha) \right), \cr
A_{2} &=& \frac{1}{2}x_{1} e^{-2Q\rho_{+}}  \left( Q_{A}\cos(\alpha) -Q_{B}\sin(\alpha) \right).\label{leo10}
\eea

Relabelling $(\theta_{A},\phi_{A},\theta_{B},\phi_{B},x,\psi_{B})\rightarrow (x_{3},x_{4},x_{5},x_{6},x_{7},x_{8})$, the 
pp wave solution becomes
\bea
L^{2} ds^{2} &=& 2du\left(dv + A_{1}dx_{1} + A_{2}dx_{2} - A dx_{4} - B dx_{6} \right)  + \delta_{ij}dx^{i}dx^{j}, \cr
L^{2} B_{2} &=& -A_{1} du\wedge dx_{1} - A_{2} du\wedge dx_{2} -A du \wedge dx_{4} -B du \wedge dx_{6} ,\cr
\Phi &=& 0\;, \;\;\; i,j=1,...,8.\label{ppwave21}
\eea
Let us now study a different geodesic and Penrose limit.
\subsubsection{Second limit: geodesic on $\psi_A,\psi_B,\phi_A,\phi_B$}

Another possible Penrose limit, still describing excitations in the IR of the dual field theory, 
is for a  null geodesic that moves on a combination of $\psi_A,\psi_B,\phi_A,\phi_B$, still at $x=0$ (or $x=x_0$, 
in general), as well as $\rho=\rho_+$, but now at $\theta_A=\theta_B=0$ (instead of $\pi/2$), and the three remaining coordinates 
(from which the combination of the null motion is taken) are also still fixed at 0. 

We first do the coordinate change (replacement)
\be
\psi_{A,B} \rightarrow  \hpsi_{A,B} - \hphi_{A,B}, \quad
        \theta_{A,B} \rightarrow 2\htheta_{A,B}, \quad
        \phi_{A,B} \rightarrow \hphi_{A,B} + \hpsi_{A,B}.
\ee

Then, we consider the same rotation as in eq.(\ref{rota1}), but now for $\hat \psi_A$ and $\hat\psi_B$ (instead of $\psi_A$ and $\psi_B$), 
\bea
\hpsi_{A} &\rightarrow& \frac{1}{\sqrt{N_{B}}} 
                \left( \cos(\alpha)\hpsi_{A} + \sin(\alpha)\hpsi_{B}\right),\cr
        \hpsi_{B} &\rightarrow& \frac{1}{\sqrt{N_{A}}}
                \left( -\sin(\alpha)\hpsi_{A} + \cos(\alpha)\hpsi_{B}\right),
\eea
such that the geodesic motion is on the redefined $\hat\psi_A$.
We use the same redefinition of lightcone coordinates $u,v$ in eq.(\ref{light1}), for $(t,\hat\psi_A)$.

We perform the coordinate change indicated in eq.(\ref{rescalingwave1}), together with the Penrose rescaling
\bea
 v&\rightarrow& \frac{v}{L^{2}}, \quad 
        \htheta_{A,B} \rightarrow \frac{\htheta_{A,B}}{L\sqrt{N_{B,A}}}, \quad
        x \rightarrow \frac{x}{L}, \quad
        \rho \rightarrow \rho_{+} + \frac{\rho}{L^2}, \quad
        \hpsi_{B} \rightarrow \frac{\hpsi_{B}}{L}\cr
u&\rightarrow& u,\quad \hat \phi_A\rightarrow \hat\phi_A,\quad \hat\phi_B\rightarrow\hat\phi_B.
\eea

Multiplying the metric by $L^2$ and taking the $L\rightarrow \infty$ limit gives 
\bea
L^{2} ds^{2} &=& -\frac{1}{2}\left( \frac{\htheta^{2}_{A}}{N_{B}} \cos^{2}(\alpha) + \frac{\htheta^{2}_{B}}{N_{A}} 
\sin^{2}(\alpha) \right)du^{2} + 2du\left(dv+  A_{\varphi}d\varphi\right)       \cr
        && + dx^{2}   + d\rho^{2} +  \rho^{2} d\varphi^{2} 
            +  d\htheta^{2}_{A} + \htheta^{2}_{A}d\hphi^{2}_{A}
            +  d\htheta^{2}_{B} + \htheta^{2}_{B}d\hphi^{2}_{B} + d\hpsi^{2}_{B},\cr
        L^{2}H_{3} &=& \cos(\alpha)\sqrt{\frac{2}{N_{B}}} \htheta_{A} du\wedge d\htheta_{A} \wedge d\hphi_{A}    
            - \sin(\alpha)\sqrt{\frac{2}{N_{A}}} \htheta_{B}du\wedge d\htheta_{B} \wedge d\hphi_{B},\cr
              && + dA_{\varphi} \wedge d\varphi\wedge du   \cr
            \Phi &=& 0\;,
\eea
where $A_{\varphi}$ is the same as in the geodesic in the previous case, eq.(\ref{diego10}). 
We can also find the 2-form potential that leads to $H_{3}=dB_2$,
    \begin{equation}
        L^{2} B_{2} = \frac{1}{\sqrt{2N_{B}}}\cos(\alpha) \htheta^{2}_{A} du \wedge d\hphi_{A} 
        - \frac{1}{\sqrt{2N_{A}}}\sin(\alpha) \htheta^{2}_{B} du \wedge d\hphi_{B} - A_{\varphi} du \wedge d\varphi.
    \end{equation}

In the above pp wave solution, we have three 2-dimensional subspaces written in polar coordinates: 
$(\rho,\varphi)$  and $(\htheta_{A,B},\hphi_{A,B})$. We switch to Cartesian coordinates $(x_{1},x_{2})$, $(x_{3},x_{4})$ and 
$(x_{5},x_{6})$, respectively. Relabelling $(x,\hpsi_{B})\rightarrow (x_{7},x_{8})$) we find
\bea
        L^{2} ds^{2} &=& -\frac{1}{2}\left( \frac{1}{N_{B}}(x^{2}_{3}+x^{2}_{4}) \cos^{2}(\alpha) + \frac{1}{N_{A}}(x^{2}_{5}+x^{2}_{6}) \sin^{2}(\alpha) \right)du^{2} \cr
        &&+ 2du\left(dv +  A_{1}dx_{1} +  A_{2}dx_{2}\right)  + \delta_{ij}dx^{i}dx^{j} ,\cr
        L^{2}B_{2} &=& - A_{1} du \wedge dx_{1} - A_{2} du \wedge dx_{2} + B_{3}du\wedge dx_{3} \cr
&&        + B_{4}du\wedge dx_{4} + B_{5}du\wedge dx_{5} + B_{6}du\wedge dx_{6} ,\cr
        \Phi &=& 0\;,   \label{ppwave12}
\eea
where $A_1,A_2$ are written in eq.(\ref{leo10}) and
    \begin{equation}
    \begin{aligned}
        B_{3} &=   \frac{1}{\sqrt{2N_{B}}}\cos(\alpha) x_{4},\quad 
        B_{4} =   -\frac{1}{\sqrt{2N_{B}}}\cos(\alpha) x_{3}, \\
        B_{5} &= -  \frac{1}{\sqrt{2N_{A}}}\sin(\alpha) x_{6}, \quad
        B_{6} =   \frac{1}{\sqrt{2N_{A}}}\sin(\alpha) x_{5}.
    \end{aligned}      
    \end{equation}

This pp wave solution is called a "gyratonic" pp wave \cite{Podolsky:2014lpa,Frolov:2005zq}, being created by spinning objects 
moving at the speed of light. Let us discuss the equivalence of the plane waves obtained above and the preserved SUSY.

\subsubsection{Equivalence of pp waves and supersymmetry}
We now show that both pp waves in eqs.(\ref{ppwave21}) and (\ref{ppwave12}) are the same, and moreover are of a general type called parallelizable 
in \cite{Sadri:2003ib}.

We start by denoting, in (\ref{ppwave21}), 
\be
2A_1=-a x_2;,\;\; 2A_2=+ax_1\;,\;\; 2A=b x_3\;,\;\;
2B=c x_5\;,
\ee
where $a,b,c$ are given by 
\be
a =  e^{-2Q\rho_{+}}  \left( Q_{A}\cos(\alpha) - Q_{B}\sin(\alpha) \right), \quad
        b = 2\sqrt{\frac{1}{2N_{B}}} \cos(\alpha), \quad 
        c = 2 \sqrt{\frac{1}{2N_{A}}} \sin(\alpha).\label{abcdef}
\ee

The pp wave solution  in eq.(\ref{ppwave21}) becomes then 
\bea
ds^2&=&2dudv +du\left[-a x_2dx_1+ax_1dx_2-2bx_3dx_4-2cx_5dx_6\right]+\sum_{i=1}^8dx_idx_i,\cr
B_{2} &=& \frac{a}{2}\; du \wedge\left( x_{2} dx_{1} - x_{1} dx_{2}\right) -b\,x_{3} du \wedge dx_{4} -c
x_{5} du \wedge dx_{6} ,\cr
\Phi &=& 0.
\eea
We can do a similar coordinate transformation as was done, for instance, in \cite{Araujo:2017hvi}, and first define
\be
z_1\equiv x_1+ix_2\;,
\ee
such that
\be
x_1dx_2-x_2dx_1=-\frac{i}{2}(\bar zdz -zd\bar z)\;, \quad dx^{2}_{1}+dx^{2}_{2} = dzd\bar{z}\;,
\ee
and then, redefining
\be
z_1=e^{-iau/2}w_1\;,\;\;
\bar z_1=e^{+iau/2}\bar w_1\;,\label{ztransf}
\ee
we find 
\be
dz_1d\bar z_1-\frac{i}{2}adu(\bar z_1dz_1-z_1d\bar z_1)=dw_1d\bar w_1-\frac{a^2}{4}du^2|w_1|^2\;,
\ee
so the metric becomes
\be
ds^2=2dudv +du\left[-\frac{a^2}{4}|w_1|^2-2bx_3dx_4-2cx_5dx_6\right]+dw_1d\bar w_1+\sum_{i=3}^8dx_idx_i.
\ee

Then, the shift
\be
v\rightarrow v +\frac{b}{2} x_3x_4+\frac{c}{2}x_5x_6\label{vtransf}
\ee
takes the metric to 
\bea
ds^2&=&2dudv +du\left[-\frac{a^2}{4}|w_1|^2du-b(x_3dx_4-x_4dx_3)-c(x_5dx_6-x_6dx_5)\right]\cr
&&+dw_1d\bar w_1
+\sum_{i=3}^8dx_idx_i.
\eea

Finally, we repeat for the pairs $(x_3,x_4)$ and $(x_5,x_6)$ the same steps as for $(x_1,x_2)$, and get
\be
ds^2=2dudv-du^2\left[\frac{a^2}{4}|w_1|^2+\frac{b^2}{4}|w_2|^2+\frac{c^2}{4}|w_3|^2\right]+\sum_{a=1}^3dw_ad\bar w_a
+dx_7^2+dx_8^2.
\ee

We go back to real coordinates by $w=x_1'+ix_2'$ and drop the primes, to find the metric 
\be
ds^2=2dudv -\frac{du^2}{4}\left[a^2(x_1^2+x_2^2)+b^2(x_3^2+x_4^2)+c^2(x_5^2+x_6^2)\right]+\sum_{i=1}^8 dx_i^2.
\ee

The field strength of the B field is, originally, 
\be
H_3=dB_2=du\wedge \left[a\, dx_1\wedge dx_2+b\, dx_3\wedge dx_4+c\, dx_5\wedge dx_6\right]\;,
\ee
but then neither the transformation (\ref{ztransf}), nor the shift (\ref{vtransf}) changes it, so it has the same form in the final 
variables. We can, moreover, choose a gauge in which
\be
B_2=\frac{1}{2}du \wedge \left[a(x_2\, dx_1-x_1\, dx_2)+b(x_4\, dx_3-x_3\, dx_4)+c(x_6\, dx_5-x_5\, dx_6)\right].
\ee

As before, $\Phi=0$.
For the pp wave on the second geodesic, see eq.(\ref{ppwave12}), using the same definitions in (\ref{abcdef}), we can put it in the form
\bea
ds^2&=& 2dudv-du[x_2dx_1-x_1dx_2]-\frac{du^2}{4}\left[b^2(x_3^2+x_4^2)+c^2(x_5^2+x_6^2)\right]+\sum_{i=1}^8 dx_i dx_i\cr
B_2&=&\frac{du}{2}\left[a(x_2\,dx_1-x_1\, dx_2)+b(x_4\, dx_3-x_3\, dx_4)+c(x_6\, dx_5-x_5\, dx_6)\right]\cr
\Phi&=&0.
\eea

We see that, with respect to the previous case, we have an intermediate case, where the transformations were done for the 
$(x_3,x_4)$ and $(x_5,x_6)$ pairs, but it remains to do for the $(x_1,x_2)$ pair. Once that is done, the same solution is obtained. 

This common solution is also of the type called "parallelizable" in \cite{Sadri:2003ib}, 
namely a solution of the type 
\bea
ds^2&=&2dudv-(du)^2\left[a_1^2(x_1^2+x_2^2)+a_2^2(x_3^2+x_4^2)+a_3^2(x_5^2+x_6^2)+a_4^2(x_7^2+x_8^2)\right]+ \sum_{i=1}^8 dx_i^2\cr
H&=&du\wedge \left(2a_1\;dx_1\wedge dx_2+2a_2\; dx_3\wedge dx_4+2a_3\; dx_5\wedge dx_6+2a_4\; dx_7\wedge 
dx_8\right)\Rightarrow\cr
B&=&du\wedge \left[a_1(x_1dx_2-x_2dx_1)+a_2(x_3dx_4-x_4dx_3)\right.\cr
&&\left.+a_3(x_5dx_6-x_6dx_5)+a_4(x_7dx_8-x_8dx_7)\right].\label{diegoarmando}
\eea
From the table on page 16 in the paper \cite{Sadri:2003ib}, we 
see that the solution (since it is of the generic type) has only 16 supercharges (a generic pp wave has 16 supercharges
or 1/2 susy, but depending on the solution, it can have perhaps more). 

On the other hand, we remember the definitions (\ref{abcdef}), as well as the fact, shown in \cite{Nunez:2023nnl}, that the 
background is supersymmetric if 
\be
e_AQ_B=\pm e_B Q_A\Rightarrow Q_B \sqrt{N_A}=\pm Q_A\sqrt{N_B}.
\ee

Then, we {\em choose} the rotation parameter $\a$ (which was free until now) such that 
\be
\cos\a\propto \sqrt{N_A}\;,\;\;
\sin\a\propto \sqrt{N_B}\Rightarrow \tan\a=\sqrt{\frac{N_B}{N_A}}.
\ee

Then we obtain $b=c$, and moreover $a=0$, for the supersymmetric background. This makes sense, since from the table 2 on page 16 of 
\cite{Sadri:2003ib}, we see that in this case 
we have 24 supercharges: the 16 generic 
ones, plus 8 additional ones. In this case we have $U(2)\times O(4)$ symmetry (R-symmetry in the field theory). 

We summarize (up to this point). We study the non-singular background corresponding to I-branes compactified on a circle and preserving SUSY in eq.(\ref{diego})--in the NS-frame. We consider its Penrose limit along two possible geodesics, showing that the result for both geodesics is the same. A parallelizable background in eq.(\ref{diegoarmando}) for $a_1=a_4=a=0$ and $a_2=a_3=b=c$,  preserving twenty-four SUSYs. Let us now study the quantization of a string on this plane wave background.

\subsubsection{Quantization of string modes}
We start by writing the equations of motion for 
a string in generic parallelizable plane wave in eq.(\ref{diegoarmando}).
For constant values of $a_i$ with $i=1,2,3,4$, and choosing the light-cone gauge $u=\alpha' p^+ \tau$ we find for the first pair of coordinates $(X^1, X^2)$,
\begin{eqnarray}
 & & \Box X^1   -\left(a_1\alpha' p^+\right)^2 X^1 + 2\left( a_1\alpha' p^+ \right)\partial_\sigma X^2=0\label{eqX1}\\
 & & \Box X^2   -\left(a_1\alpha' p^+\right)^2 X^2 - 2\left( a_1\alpha' p^+ \right)\partial_\sigma X^1=0.\nonumber
\end{eqnarray}
There are  analog equations for the pairs $(X^3,X^4)$ with the constant $a_2$. For $(X^5,X^6)$ with constant $a_3$ and for $(X^7,X^8)$ with constant $a_4$. The equation for the $u$-coordinate is implied by the Virasoro constraint and the equation for the $v$-coordinate is automatically satisfied. 

In the case of our plane wave with 24 SUSYs ($a=0$, $b=c$), the string spectrum on the pp wave is easily obtained. 
In  the light-cone gauge $\d_\tau u =\a'p^+$, and with the usual ansatz
\be
X^1=X_0^1 e^{-i\omega t+in \sigma}\;,\;\;
X^2=X_0^2 e^{-i\omega t+ in \sigma}\;,
\ee
the equations of motion for the (decoupled from the rest) $(x_1,x_2)$ system become 
\bea
 (\omega^2-n^2-(a\a'p^+)^2)X_0^1 +2in (a\a'p^+)X_0^2&=&0\cr
(\omega^2-n^2-(a\a'p^+)^2)X_0^2-2in (a\a'p^+)X_0^1&=&0\;,
\eea
so that
\bea
\frac{X_0^2}{X_0^1}&=&\frac{+2in (a\a'p^+)}{\omega^2-n^2-(a\a'p^+)^2}=\frac{\omega^2-n^2-(a\a'p^+)^2}{-in(a\a'p^+)}
\Rightarrow\cr
\omega&=&\sqrt{n^2+(a\a'p^+)^2\pm 2n (a\a'p^+)}=|n\pm a\a'p^+|\;,
\eea
but, with the usual rescaling of $\sigma$, this gives 
\be
\omega=\left|a\pm \frac{n}{\a'p^+}\right|.
\ee
The same for  the coordinates $(x_3,x_4)$ and $(x_5,x_6)$ and constants $b, c$.
In the SUSY case ($a=0,~ b=c$), we have 
\be
\omega_\pm =\left|b\pm \frac{n}{\a'p^+}\right|\;, i=3,4,5,6\;,\;\;
\omega=\frac{n}{\a'p^+}\;,\;\; i=1,2, 7,8.
\ee
so two complex modes of the same mass $b$, and two complex massless modes since, coming from the $(x_1,x_2)$ and $(x_7,x_8)$ pairs.

\subsection{Dual spin chain}

Now we would like to try to reproduce the string spectrum on the pp wave derived above from the dual 
quantum field theory, originally defined in 1+1 dimensions. Here, we only concentrate on the $n=0$ modes, 
corresponding to the BPS operators in the supersymmetric field theory case. The $n/(\a' p^+)\propto 1/J$ 
terms are  left for future work.

\subsubsection{Field theory}

As explained in \cite{Itzhaki:2005tu} and \cite{Nunez:2023nnl}, the $(1+1)-$dimensional theory, with coordinates $x_0$ and 
$x_1$, comes from the intersection of two sets of D5-branes in type IIB string theory. The corresponding low-energy field theory is 
the theory with gauge groups $SU(N_A)_{N_B}\times SU(N_B)_{N_A}$, with both YM and CS terms in $(2+1)$ dimensions, 
and with fermion bifundamentals under the gauge groups, reduced to $(1+1)$ dimensions. The theory also has an 
$SO(4)\times SO(4)$ R-symmetry, corresponding to the two three-spheres $S^3$'s in the gravity dual, see eq.(\ref{diego}). This is manifest in field theory as the rotation of the 
two sets of four coordinates parallel to each D5-brane, but transverse to the $(1+1)-$dimensional intersection.

The $(2+1)-$dimensional theory (before the reduction) has some similarity with the theory for the 2+1 dimensional GJV model
\cite{Guarino:2015jca} (which is also of  YM + CS type, though only for one gauge group), whose spin chain coming from a Penrose
limit was described in \cite{Araujo:2017hvi}. It is therefore conceivable that one could write the action, in a similar way with the GJV 
case, then reduce on the $S^1$. Parts of the theory were written down 
 in \cite{Itzhaki:2005tu}. The authors of that paper considered only the gauge fields $A_{(1)}$ and 
$A_{(2)}$, coming from the two sets of D5-branes, and their interaction with the bifundamental fermions. One should  
have
the scalars of the D5-branes, 
$\phi^I_{(1)}$, with $I=2,3,4,5$ (transverse to the first set of D5-branes) and $\phi^J_{(2)}$ (transverse to the second set of D5-branes), 
with $J=6,7,8,9$, on which the $SO(4)\times SO(4)$ R-symmetry should act.

In the Penrose limit, the 8 scalars will pair up into 4 complex ones, $Z,\bar Z$ (corresponding to the $u$ and $v$ 
directions in the gravity dual), and $W_1=X_1+iX_2, W_2=X_3+iX_4,W_3=X_5+iX_6$ (we renamed the 
scalars, without explaining the notation, since we don't yet know how they generically pair up into these). Then the spin chain 
will be, as usual, with insertions of $W_i, \bar W_i$ (6 oscillators) and $D_i$ (2 oscillators) inside the trace of the operator 
corresponding to the vacuum, $\Tr[Z^J]$.

It is not clear how to get the case of frequencies of string oscillators with arbitrary $a,b,c$'s from field theory, though perhaps 
that is because of having less supersymmetries, and so there will be more arbitrary renormalizations when going from 
weak coupling (SYM) to strong coupling (AdS). 

However, the supersymmetric case is easier to understand. In this case, we construct the complex scalars
\be
Z=\phi_{(1)}^2+i\phi_{(2)}^6\;,\;\;
W_1=\phi_{(1)}^3+i\phi_{(2)}^7\;,\;\;
W_2=\phi_{(1)}^4+i\phi_{(1)}^5\;,\;\;
W_3=\phi_{(2)}^8+i\phi_{(2)}^9.
\ee

Then, from the $U(2)\times O(4)$ R-symmetry derived from the supersymmetric pp wave previously found,
the $SO(4)$ acts on $W_2,W_3$ (that have the same value for $\Delta-J$, corresponding to the same mass, $b$,
in the supersymmetric pp wave), and the $U(2)$ acts on $(Z,W_1)$ (both corresponding to massless fields in the 
supersymmetric pp wave). 
There are also the modes corresponding to insertions of $D_i$, $i=0,1$, but those correspond to the directions parallel to the 
field theory, so they are special. 

\subsubsection{Matching of oscillator modes and masses: naive try}

The correspondence between the pp wave (and gravity dual) coordinates and scalar fields in field theory is then as follows:
\bea
&&(x_1,x_2)=(\rho,\varphi)\rightarrow D_i\;,\;\;\; (x_7,x_8)=(x,\psi_B)\rightarrow W_1\;,\cr
&&(x_3,x_4)=(\theta_A,\phi_A)\rightarrow W_2\;,\;\;\; (x_5,x_6)=(\theta_B,\phi_B)\rightarrow W_3\;.
\eea

The first, {\em naive}, try for the oscillator spectrum is as follows (we are in 1+1 dimensions, so a scalar has mass 
dimension 0)
\begin{center}
    \begin{tabular}{|c|c|c|c|c|c|c|c|}
    \hline
    field & $Z$ & $W_1$ & $\bar Z$ & $\bar W_1$ & $W_2,\bar W_2$ & $W_3,\bar W_3$ & $D_{x_i}$ \\
    \hline\hline
    $\Delta$ & 0 & 0 & 0 & 0 & 0 & 0 & 1 \\
    \hline
    $J$ & -1 & -1 & 1 & 1 & 0 & 0 & 0 \\
    \hline
    $\Delta-J$ & 1 & 1 & -1 & -1 & 0 & 0 & 1\\
    \hline
    $H/\mu=\Delta-J-E_0$ & 0 & 0 & -2 & -2 & -1 & -1 & 0\\
    \hline
    oscillator & - & $x_8$ & - & - & $x_3,x_4$ & $x_5,x_6$ & $x_1,x_2$\\
    \hline
    \end{tabular}
\end{center}

Then the Hamiltonian has the correct value, if one multiplies it by $\mu=-1$. $\bar Z$ and $\bar W_1$ would get infinite masses by 
renormalization, and thus do not correspond to oscillators, as usual. Note that $W_1$ is massless, because of having the same 
$J$ as $Z$, since $\psi_A$ (for $Z$)
and $\psi_B$ (for $W_1$) are rotated into each other. 

But we see that we are missing $x_7$ (one massless oscillator), 
and the problem is traced to the fact that $\bar W_1$ is not an oscillator, just like $\bar Z$, so it does not enter the counting. 
How to fix this? 

\subsubsection{Matching of oscillators modes and masses: correct version}

The problem is resolved if we think better of the construction in \cite{Itzhaki:2005tu}. The field theory is actually 
$(2+1)-$dimensional, where the extra spatial coordinate for the field theory 
was obtained by joining at $\bar u=\bar v=0$ the radial coordinates 
$\bar u\equiv \sqrt{\sum_i \phi_{(1)}^i\phi_{(1)}^i}$ and $\bar v=\sqrt{\sum_j\phi_{(2)}^j\phi_{(2)}^j}$ (the radial coordinates 
transverse to each D5-brane). Let us call this direction $w$. Then we must add $x_7$ (corresponding to $x$) as the 
$D_w$ oscillator in the $(2+1)-$dimensional field theory. 

Then the correct table is (note that in $(2+1)$ dimensions, a scalar has dimension 1/2, so we must redefine the unit of $J$ to be 1/2 as well)

\begin{center}
    \begin{tabular}{|c|c|c|c|c|c|c|c|c|}
    \hline
    field & $Z$ & $W_1$ & $\bar Z$ & $\bar W_1$ & $W_2,\bar W_2$ & $W_3,\bar W_3$ & $D_{x_i}$ & $D_w$\\
    \hline\hline
    $\Delta$ & 1/2 & 1/2 & 1/2 & 1/2 & 1/2 & 1/2 & 1 &1 \\
    \hline
    $J$ & -1/2 & -1/2 & 1/2 & 1/2 & 0 & 0 & 0 & 0 \\
    \hline
    $\Delta-J$ & 1 & 1 & 0 & 0 & 1/2 & 1/2 & 1 & 1\\
    \hline
    $H/\mu=\Delta-J-E_0$ & 0 & 0 & -1 & -1 & -1/2 & -1/2 & 0 & 0\\
    \hline
    oscillator & - & $x_8$ & - & - & $x_3,x_4$ & $x_5,x_6$ & $x_1,x_2$ & $x_7$\\
    \hline
    \end{tabular}
\end{center}

Now the Hamiltonian is the correct one if we multiply by $\mu=-2$. Of course, there seems to be a sort of over-counting 
the dimensions (which now sum up to 11, instead of the 10 of type IIB), but that is due to the fact that the 
dimension $w$ is half of one dimension, and half of another, so does not truly exist independently: any point with 
$(\bar u\neq 0,\bar v\neq 0)$, does not belong to the worldvolume of the QFT, and the fibered I-brane gravity dual we 
consider only deals with these points. 

In any case, as usual, the vacuum is given by 
\be
|0;p^+\rangle=\frac{1}{\sqrt{J} N^{J/2}}\Tr[Z^J]\;,
\ee
and we insert the oscillators from the table above inside the trace, in order to obtain the oscillator states. 

Now, we perform a similar study for the background dual to a $(4+1)$-dimensional confining QFT \cite{Nunez:2023xgl}.
\section{Penrose limit of twisted D5-branes and dual spin chain}\label{sectiontwistedD5}
In \cite{Nunez:2023xgl}, a gravity dual was considered for a single set of twisted compactified D5-branes on $S^1_\varphi$, with a worldvolume of $(t,\vec{x}_4)$, 
plus a  fibration over the coordinate $\varphi$: the $S^3$ transverse to the single 
set of D5-branes is fibered over $\varphi$, obtaining a cigar-like geometry. 
One has a Wilson line holonomy inserted, and the theory preserves 8 supercharges (1/4 susy). 

The gravity dual background, in the NS5-brane frame, is given by 
\bea
ds^{2}_{st} &=&  dx^{2}_{1,4} + f_{s}(\rho)d\varphi^{2} 
            + \frac{d\rho^{2}}{f_{s}(\rho)} 
            + \frac{N}{4}\left(\omega^{2}_{1} + \omega^{2}_{2} + 
        \left(\omega_{3} - \sqrt{\frac{8}{N}} Q \zeta(\rho) d\varphi\right)^{2}\right) ,\cr
        H_{3} &=& 2N \Vol(S^{3}) +2 \sqrt{\frac{N}{8}}Q d\left( \zeta(\rho) \omega_{3}\wedge d\varphi \right),\cr
        \Phi &=& - \frac{\rho}{\sqrt{N}}\;,
\eea
where \footnote{Note that here $e^{2\rho_-/\sqrt{N}}<0$, due to the fact that $Q^2$ has the opposite sign to the one in a black solution. That is so, since the solution is related to the black membrane through a double Wick rotation, that requires also $Q\rightarrow -iQ$ in order to keep the solution real. But the negative sign is no problem: the double Wick rotation means that $\rho_\pm$ are not interpreted as horizons: only $\rho_+$ is physical, and it is just the tip of a cigar geometry, locally the origin of a tangent plane. $\rho_-$ on the other hand is just a mathematical construct.  }
\bea
f_{s}(\rho) &=& 1 - m e^{-2\rho/\sqrt{N}}- 2Q^{2} e^{-4\rho/\sqrt{N}} \cr
&=& e^{-4\rho/\sqrt{N}}\left(e^{2\rho/\sqrt{N}}
-e^{2\rho_{+}/\sqrt{N}}\right)\left(e^{2\rho/\sqrt{N}}-e^{2\rho_{-}/\sqrt{N}}\right),\cr
        \zeta(\rho) &=& e^{-2\rho/\sqrt{N}} - e^{-2\rho_{+}/\sqrt{N}}\;,
\eea
and, as in the I-brane case, the metric on the transverse $S^3$ is written in terms of the Maurer-Cartan forms of $su(2)$, 
\bea
\omega_{1} &=& \cos(\psi)d\theta +\sin(\psi)\sin(\theta)d\phi,\cr
        \omega_{2} &=& -\sin(\psi)d\theta) +\cos(\psi)\sin(\theta)d\phi,\cr
        \omega_{3} &=& d\psi +\cos(\theta_A)d\phi.
\eea

The $\rho_\pm$ are the two solutions of $f_s(\rho)=0$, namely 
\be
{e^{2\frac{\rho_\pm}{\sqrt{N}}}=\frac{m\pm \sqrt{m^2+8Q^2}}{2}.}
\ee

The supersymmetric case corresponds to 
\be
e^{\frac{2\rho_+}{\sqrt{N}}}=\sqrt{2} Q,~~~~m=0.
\ee

As in the I-brane case, the end of the cigar, $\rho=\rho_+$, corresponds to the IR of the field theory.

\subsection{Penrose limit}

In order to explore the IR of the field theory, we consider a geodesic at $\rho=\rho_+$, like in the I-brane case of Section \ref{sectionIbranes}. 
Moreover, we now consider a geodesic moving in $t$ and $\psi$ (the coordinate on an equator of $S^3$) and at $\theta=\pi/2, \phi=0$, 
as well as $x^i=0$, $i=1,...,4$ (or any constant $x^i_0$), and $\varphi$ arbitrary. 

We define the usual lightcone coordinates, now for $(t,\psi)$, 
\bea
t &=& \frac{1}{\sqrt{2}}(u-v),\cr
\psi&=& \frac{1}{\sqrt{2}}(u+v),
\eea
and then the standard rescalings in the Penrose theorem, 
\bea
v&\rightarrow& \frac{v}{L^{2}}, \quad 
        \theta \rightarrow \frac{\pi}{2}+\frac{2\theta}{L\sqrt{N}}, \quad
        \phi \rightarrow \frac{2\phi_{A,B}}{L\sqrt{N}}, \quad
        x^{i} \rightarrow \frac{x^{i}}{L}, \quad
        \rho \rightarrow \rho_{+} + \frac{\rho}{L^2},\cr
u&\rightarrow & u,\quad \varphi\rightarrow\varphi\;,
\eea
where the only nontrivial case is that of $\varphi$, which does not rescale, since it is an angular coordinate (and stays so). 

Then we take the Penrose limit, by multiplying the metric with $L^2$ and taking the limit $L\rightarrow \infty$, 
while also using the redefinition
\be
\rho \rightarrow \frac{1}{\sqrt{N}} \frac{\sinh\left( \frac{1}{\sqrt{N}}(\rho_{+}-\rho_{-})\right)}{e^{\frac{1}{\sqrt{N}}
(\rho_{+}-\rho_{-})}} \rho^{2}, \quad
\varphi \rightarrow \frac{\sqrt{N}}{2}\frac{e^{\frac{1}{\sqrt{N}}(\rho_{+}-\rho_{-})}}{4 \sinh\left(\frac{1}{\sqrt{N}}(\rho_{+}-\rho_{-})\right)} 
\varphi\;,
\ee
to obtain the pp wave solution 
\bea
L^{2}ds^{2} &=& 2 du \left( dv - \sqrt{\frac{2}{N}} \theta d\phi
                 + \frac{Q}{\sqrt{N}}e^{-2\rho_{+}/\sqrt{N}}\rho^{2} d\varphi\right)^{2} + d\rho^{2} + \rho^{2}d\varphi^{2}
                 + d\theta^{2} + d\phi^{2} + d\vec{x}^{2},\cr
L^{2}H_{3} &=&  \sqrt{\frac{2}{N}} du \wedge d\theta \wedge d\phi
                + \frac{2Q}{\sqrt{N}} e^{-2\rho_{+}/\sqrt{N}} \rho du \wedge d\rho \wedge d\phi,\cr       
\Phi &=& 0.
\eea
We now move to Cartesian coordinates for $(\rho,\varphi)$, namely $(x^5,x^6)$, and relabel the rescaled $(\theta,\phi)$ as 
$(x^7,x^8)$, to obtain 
\bea
ds^{2}_{PL} &=& 2 du \left( dv - \sqrt{\frac{2}{N}} x_{7} dx_{8}
                 + \frac{Q}{\sqrt{N}}e^{-2\rho_{+}/\sqrt{N}}(-x_{6}dx_{5}+x_{5}dx_{6})\right)^{2} + d\vec{x}^{2},\cr
        H_{3} &=&  \sqrt{\frac{2}{N}} du \wedge dx_{7} \wedge dx_{8}
                + \frac{2Q}{\sqrt{N}} e^{-2\rho_{+}/\sqrt{N}} du \wedge dx_{5}\wedge dx_{6},\cr        
        \Phi &=& 0.
\eea

Defining 
\be
a = \frac{Q}{\sqrt{N}}e^{-2\rho_{+}/\sqrt{N}}, \quad
        b = \frac{1}{2}\sqrt{\frac{2}{N}}\;,\label{mario10}
\ee
we obtain the solution in the form
\bea
ds^{2}_{PL} &=& 2 du \left[dv  +a (- x_{6}dx_{5} + x_{5}dx_{6})  
                    - 2b x_{7} dx_{8} \right] + d\vec{x}^{2},\cr
        H_{3} &=&  2a du \wedge dx_{5}\wedge dx_{6} 
                +2b du \wedge dx_{7} \wedge dx_{8} ,\cr        
        \Phi &=& 0.
\eea

\subsubsection{Coordinate change to parallelizable pp wave, and supersymmetry}
We make a coordinate change to a parallelizable pp wave. Indeed, defining the complex coordinate
\be
z=x_5+ix_6\;,
\ee
we get
\be
-x_{6}dx_{5}+x_{6}dx_{5} = -\frac{i}{2}( \zb dz - z d\zb).
\ee

Then, the coordinate change 
\be
z = e^{-i a u/2}w, \quad \zb = e^{i a u/2} \wb
\ee
means that 
\be
dzd\zb + 2du\left( -\frac{ia}{2}( \zb dz - z d\zb   ) \right) = dwd\wb - a^{2} w\wb du^{2}\;.
\ee
Going back to Cartesian coordinates, we get the pp wave solution in the form
\bea
ds^{2} &=& -a^{2}(x^{2}_{5} + x^{2}_{6})du^{2}
                        + 2 du \left( dv - 2b x_{7} dx_{8} \right) 
                            + d\vec{x}^{2},\cr
        H_{3} &=&  2a du \wedge dx_{5}\wedge dx_{6} 
                +2b du \wedge dx_{7} \wedge dx_{8} ,\cr   
        \Phi &=& 0.
\eea

Under these coordinate changes, $H_3$ did not change, as in the I-brane case.
Finally, we shift $v$ as
\be
v \rightarrow v + b x_{7}x_{8},
~~~~dv \rightarrow dv + b x_{8} dx_{7}                 + b x_{7} dx_{8}\;,
\ee
which means that we put the term proportional to $b$ in the same form as the term proportional to $a$ was, so we can follow 
the previous steps again for this term. 

Finally, the result is that the pp wave is put in the parallelizable form
\bea
ds^{2} &=& 2dudv-\left[a^{2}(x^{2}_{5} + x^{2}_{6}) 
                        + b^{2}(x^{2}_{7} + x^{2}_{8}) \right]du^{2}
                          + d\vec{x}^{2},\cr
        H_{3} &=&  2a du \wedge dx_{5}\wedge dx_{6} 
                +2b du \wedge dx_{7} \wedge dx_{8} ,\cr       
        \Phi &=& 0,
\eea
with $a,b$ defined in eq.(\ref{mario10}).
We also observe that the supersymmetric case, which is now also the extremal case, with $a=b$, 
\be
e^{\frac{2\rho_+}{\sqrt{N}}}=\sqrt{2}Q,~~~~m=0,
\ee
has the same form as the supersymmetric pp wave obtained from the fibered I-branes, 
which in the classification in \cite{Sadri:2003ib}, 
has 24 supercharges: the 16 generic ones, plus additional 8 ones.
The R-symmetry is also the same as before, $U(2)\times O(4)$.

\subsection{Dual spin chain}

The natural field theory dual would be the theory on the worldvolume of the D5-branes, with $(t,\vec{x}_4)$ and the compact 
$\varphi$, on which we can compactify. As we said, the theory has implicitly a Wilson loop holonomy, breaking a quarter of the SUSY, preserving eight supercharges. 

As in the I-brane case, we will only be able to describe the supersymmetric case, corresponding in the gravity dual to 
$a=b$. The generic case is unclear, but we expect that quantum corrections would spoil that analysis in any case. 

We will see that, in fact, it is better to consider the same $(1+1)-$dimensional field theory as for the I-brane case 
(also since the supersymmetric pp wave is the same), but now understood as coming from the compactification of the 
$(4+1)-$dimensional theory in $(t,\vec{x}_4)$ on the missing $S^3$ (which was transverse in the I-brane case, but is now parallel 
in this fibered D5-brane case). 

Then the field theory scalars are 4 $X$'s transverse to the D5-branes, let us call them $X^6, X^7, X^8, X^9$, plus 
4 for the compactification on $S^3\times S^1$, let's call them 
%
$A_2,A_3,A_4,A_5$, 
as they arise from gauge fields in compact directions that become scalars after compactification. 

Then, if $\psi$ rotates $X^6, X^7$, we take the complex scalar field $Z=X^6+iX^7$ to be charged under $J$. 
As in the I-brane case, this has $\Delta=0, J=-1$, so gives $\Delta-J=1$, just like $D_i$, $i=0,1$, which now has $\Delta=1, J=0$, 
and after subtracting the ground state energy 
$E_0=1$, gives 2 of the massless oscillators. Then $X^8, X^9$ have $\Delta=0, J=0, \Delta-J=0$, 
so after subtracting $E_0=1$ gives $-1$, so we obtain 2 of the massive oscillators: $(x_7,x_8)=(\theta,\phi)$ in the gravity dual. 

However, in order to obtain the correct number of massless and massive oscillators, we now
need to split $A_a$, where $a=2,3,4,5$ (for the 4 coordinates on $S^3\times S^1$) into 
2 massless oscillators and 2 massive oscillators (for $(x_5,x_6)=(\rho,\varphi)$ in the gravity dual).

To do so, we can write: $A_5\rightarrow X_\varphi$, so the field in the $\varphi$ direction is thought of as a scalar: 
by multiplication with $1/g^2_{\rm YM}$, it has dimension 0 instead of dimension 1. Moreover, we can do the same for the 
overall field on the $S^3$, 
\be
\frac{1}{g^2_{\rm YM}}\sqrt{\sum_{a=2}^4 A_a A_a}\rightarrow X_\rho\;,
\ee
with dimension 0, like a scalar. Then $X_\varphi, X_\rho$ have $\Delta-J=1$, and also correspond to massless excitations, 
specifically $x_5,x_6$ (for $\varphi,\rho$). 
This leaves $A_2, A_3$ as dimension 1 fields, giving the remaining 2 massless modes. 

This procedure is consistent, but somewhat puzzling, yet is the only one possible that gives the same result as in the I-brane 
case, which was necessary, since the supersymmetric pp wave was the same.

As for the other fields, we shift the mass by the ground state energy $E_0=1$, and multiply it by $\mu=-1$. 

Thus the final table is 
\begin{center}
    \begin{tabular}{|c|c|c|c|c|c|c|c|}
    \hline
    field & $Z$ & $\bar Z$ & $X^3,X^4$ & $X_\varphi$ & $X_\rho$ & $A_{a'}, a'=2,3$ & $D_{x_i}, i=0,1$ \\
    \hline\hline
    $\Delta$ & 0 & 0 & 0 & 0 & 0 & 1 & 1 \\
    \hline
    $J$ & -1 & 1 & 0 & 0 & 0 & 0 & 0 \\
    \hline
    $\Delta-J$ & 1 & -1 & 0 & 0 & 0 & 1 & 1\\
    \hline
    $H/\mu=\Delta-J-E_0$ & 0 & -2 & -1& -1 & -1 & 0 & 0\\
    \hline
    oscillator & - & - & $x_7,x_8$ & $x_6$ & $x_5$ & $x_3,x_4$ & $x_1,x_2$\\
    \hline
    \end{tabular}
\end{center}

The R-symmetry of $U(2)\times O(4)\simeq U(1)\times SO(3)\times SO(4)$ (modulo some discrete symmetries) in the supersymmetric case
is understood in the field theory as the symmetry of $S^1_\varphi$, $S^2\subset S^3$ (since $S^3$ was understood as an $S^2$ 
fibration, allowing for the split of $A_2,A_3,A_4$ into two massive and one massless excitation), and of $X^6,...,X^9$, respectively.

\section{Discussion and conclusions}\label{discusion-conclusion}

In this paper we have studied the Penrose limits of the gravity duals of fibered I-branes and D5-branes 
\cite{Nunez:2023nnl,Nunez:2023xgl}, in order to understand better the duality. 

The I-branes gave a $(1+1)-$dimensional theory, dynamically extended to $(2+1)$ dimensions.
We have found that in order to match the pp wave analysis obtained in the Penrose limit against a
field theory, we must consider the theory in $(2+1)$ dimensions, in which case we can match the oscillators and their masses.

Perhaps more surprisingly, we have found that, in order to match the oscillators and masses for the case of the 
Penrose limit of the twisted $S^1$-compactified  D5-branes against field theory, we need to consider the same $(1+1)-$dimensional theory 
of the original I-brane worldvolume, now obtained by reduction on the $S^1$ fibration circle, as well as the worldvolume
$S^3$. Moreover, in the $S^3$, one of the directions becomes special also, although our analysis still is made from 
the $(1+1)-$dimensional point of view, and not a $(2+1)-$dimensional one. 

In both fibered I-branes and D5-branes cases, we find that the {\em supersymmetric} pp wave, the only 
case for which we can find matching, is the same one, the generic parallelizable 
case with 24 supercharges (the standard 16 SUSYs, plus 8 more) from the classification in \cite{Sadri:2003ib}. 
The non-supersymmetric pp waves are different in the two cases, but it is not clear if or how to match, as in that case, 
there will presumably be uncontrollable corrections to the mass.
We also found that, before a coordinate transformation, one of the generic Penrose limits on the I-brane case 
(with parameters $a\neq b\neq c$)
gave us a so-called gyratonic pp wave \cite{Podolsky:2014lpa,Frolov:2005zq}.

There are many issues left for further work, among them notably the match in the generic, non-supersymmetric case, as well 
as the match of the terms with $n/(\a' p^+)$  in the string oscillator frequencies, which correspond to $\propto 1/J$ 
terms, coming from gauge interactions, in the field theory side. Apart from this, backgrounds similar to the ones studied here, see for example \cite{Fatemiabhari:2024aua},\cite{Anabalon:2022aig}, \cite{Anabalon:2024qhf}, \cite{Anabalon:2024che}, \cite{Chatzis:2024top} could be analysed following our formalism. We expect results with similar qualitative features.

\section*{Acknowledgements}

The work of HN is supported in part by  CNPq grant 301491/2019-4 and FAPESP grant 2019/21281-4. 
HN would also like to thank the ICTP-SAIFR for their support through FAPESP grant 2016/01343-7. CN and RS are supported by grants  ST/Y509644-1 and ST/X000648/1. MB is supported by FAPESP grant 2022/05152-2.

\bibliography{PenroseID5branes}
\bibliographystyle{utphys}

\end{document}